\documentclass[conference]{IEEEtran}
\IEEEoverridecommandlockouts

\usepackage{algorithm}
\usepackage{algpseudocode}
\usepackage{amsmath,amssymb,amsfonts}
\usepackage{booktabs}
\usepackage{tablefootnote}
\usepackage{textcomp}
\usepackage{graphicx}
\usepackage{svg}
\usepackage{subcaption}
\usepackage{ifthen}   
\usepackage{xcolor}   
\usepackage[hidelinks]{hyperref} 

\graphicspath{{figures/}}

\newcommand{\final}{1} 
\newcommand{\muhammad}   [1]{{{\color{red}(muhammad) #1}}}
\ifthenelse{\equal{\final}{1}}{\renewcommand{\muhammad}[1]{}}{}
\newcommand{\ryan}   [1]{{{\color{orange}(ryan) #1}}}
\ifthenelse{\equal{\final}{1}}{\renewcommand{\ryan}[1]{}}{}
\newcommand{\xiaohu}   [1]{{{\color{yellow}(xiaohu) #1}}}
\ifthenelse{\equal{\final}{1}}{\renewcommand{\xiaohu}[1]{}}{}
\newcommand{\bryant}   [1]{{{\color{teal}(bryant) #1}}}
\ifthenelse{\equal{\final}{1}}{\renewcommand{\bryant}[1]{}}{}
\newcommand{\lixun}   [1]{{{\color{cyan}(lixun) #1}}}
\ifthenelse{\equal{\final}{1}}{\renewcommand{\lixun}[1]{}}{}
\newcommand{\alex}   [1]{{{\color{blue}(alex) #1}}}
\ifthenelse{\equal{\final}{1}}{\renewcommand{\alex}[1]{}}{}
\newcommand{\ganesh}   [1]{{{\color{violet}(ganesh) #1}}}
\ifthenelse{\equal{\final}{1}}{\renewcommand{\ganesh}[1]{}}{}

\newcommand{\framework}[1]{{{\textit{#1}}}}

\def\BibTeX{{\rm B\kern-.05em{\sc i\kern-.025em b}\kern-.08em
    T\kern-.1667em\lower.7ex\hbox{E}\kern-.125emX}}

\begin{document}

\title{tritonBLAS: Triton-based Analytical\\ Approach for GEMM Kernel Parameter Selection}

\author{%
  \IEEEauthorblockN{%
    Ryan Swann,
    Muhammad Osama,
    Xiaohu Guo,
    Bryant Nelson,
    Lixun Zhang,\\
    Alex Brown,
    Yen Ong,
    Ali Yazdani,
    Sean Siddens,
    Ganesh Dasika,
    Alex Underwood
  }%
  \IEEEauthorblockA{%
    Advanced Micro Devices, Inc.\\
    \texttt{\{ryan.swann, muhammad.osama, xiaohu.guo, bryant.nelson, lixun.zhang,}\\
    \texttt{alex.brown, yen.ong, ali.yazdani, sean.siddens, ganesh.dasika,}\\
    \texttt{alexander.underwood\}@amd.com}%
}}

\maketitle

\begin{abstract}
We present \framework{tritonBLAS}, a fast and deterministic analytical model that uses architectural parameters like the cache hierarchy, and relative code and data placement to generate performant GPU GEMM kernels. \framework{tritonBLAS} explicitly models the relationship between architectural topology, matrix shapes, and algorithmic blocking behavior to predict near-optimal configurations without runtime autotuning. Based on this model, we developed and implemented a lightweight GEMM framework entirely within \framework{Triton}.

We evaluate the performance of \framework{tritonBLAS} across a diverse set of GEMM problem sizes on modern GPUs.~\framework{tritonBLAS} achieves over 95\% of the performance of 
autotuning solutions, while reducing autotuning time to \textit{zero}. This makes \framework{tritonBLAS} a practical drop-in replacement for empirical tuning in production HPC and ML workloads.
\end{abstract}




\section{Introduction}
General matrix multiplication (GEMM) drives performance in AI, ML, and HPC workloads~\cite{Asanovic:2006:Dwarfs,Dongarra:1990:BLAS}. Large language models (LLMs) like GPT-4 depend on transformer networks that execute thousands to millions of GEMM operations per pass~\cite{OpenAI:2023:GPT4}. GPUs, equipped with specialized matrix cores, dominate GEMM acceleration. Despite the algorithm’s apparent simplicity, developing efficient GEMM kernels and libraries for GPUs remains a difficult task. Writing a fast GEMM kernel requires carefully capturing the GPU’s compute and memory hierarchies, orchestrating data movement across registers, shared memory, and caches, and ensuring that matrix cores are continuously supplied with data to avoid stalls.

Implementing a fast kernel is only one piece of the problem. Building a complete library requires not only a collection of optimized kernels but also the ability to select or Just-in-Time (JIT)\footnote{Just-in-Time (JIT) compilation refers to the strategy of generating and compiling specialized machine code at runtime, based on problem parameters provided by the user. This allows the library to tailor kernel implementations to the specific matrix shapes and hardware characteristics without relying on precompiled variants.} compile the right kernel configuration for the user’s problem at runtime. Such configuration decisions include the choice of tile sizes, unroll depths, loop scheduling strategies, and data reordering to improve L2 cache hit rates. The effectiveness of the library therefore depends as much on kernel selection and orchestration as it does on the raw efficiency of individual kernels.

Triton~\cite{Tillet:2019:TAI} has emerged as a productive programming model for writing high-performance GPU kernels. It simplifies much of the complexity of GPU programming such as thread synchronization, vectorization, and memory coalescing, while still exposing the necessary control for performance tuning. Triton therefore makes writing GEMM kernels significantly easier. However, Triton and similar frameworks still depend on autotuning to select kernel parameters. These search-based techniques require extensive compilation and benchmarking, consume large amounts of time and resources, and limit portability of tuned configurations across different hardware generations (see Section~\ref{sec:triton}).

We present \framework{tritonBLAS}, a first-class GEMM library built on top of Triton that eliminates reliance on autotuning. Instead of empirical search, \framework{tritonBLAS} uses an analytical performance model that captures the interaction between GPU architecture and algorithmic blocking choices. This model allows us to predict near-optimal configurations deterministically and to generate specialized kernels through JIT compilation without running a tuning phase. By combining Triton’s programmability with a principled analytical model, we provide a practical alternative to autotuning for production HPC and ML workloads. 

The contributions of this paper are as follows:
\begin{itemize}
    \item \textbf{A deterministic, zero-autotuning selector for Triton GEMM.}  
    Using this model, we implement \framework{tritonBLAS}, a Triton-based GEMM library that replaces empirical autotuning architecture-parameterized configuration selection.
    \item \textbf{Architecture portability with lightweight calibration.}  
    The model is parameterized only by measurable hardware rates (bandwidths, instruction latencies, and matrix-core shapes), enabling retargeting to new GPU generations via microbenchmark-based calibration.
    \item \textbf{Extensive evaluation demonstrating near-autotuned performance.}  
    Across 150{,}000 GEMM shapes and real-world LLM workloads, \framework{tritonBLAS} achieves $94.7\%$ of exhaustive autotuning performance while eliminating tuning overhead and matching or exceeding vendor libraries in memory-bound regimes.
\end{itemize}

\section{Background and Related Work}
\subsection{General Matrix Multiplication (GEMM)}
General matrix multiplication (GEMM) is defined as $C = \alpha AB + \beta C$ where $C$ is the output matrix of size $M \times N$, A and B are input matrices of size $M \times K$ and $K \times N$ respectively, and $\alpha$ and $\beta$ are scalar values.
Efficient GEMM implementations on GPUs take advantage of a GPU's memory and compute hierarchy through hierarchical tiling structures in attempts to maximize the balance of maximum locality and utilization of available parallelism~\cite{Kerr:2017:CUTLASS,Osama:2023:SWP}. GEMM tiles are scheduled using an ``output stationary'' dataflow with the reduction dimension, K, accumulated into an output buffer over time, and output tiles parallelized across processing elements (see Algorithm~\ref{alg:matmul}~and~\ref{alg:tiledMM}).
GEMM's output tile dimensions influence both parallelism and data locality, and are impacted by the architectural characteristics such as memory hierarchy bandwidth and capacity, as well as the shape and throughput of the compute resources.
This implies that any problem size and shape (M, N, K) of a GEMM may require a different tile size to reach peak performance on a specific architecture.

\begin{algorithm}
  \caption{Compute matrix $C = A \times B$}\label{alg:matmul}
  \begin{algorithmic}[1]
    \Require $A\in\mathbb{R}^{m\times k},\;B\in\mathbb{R}^{k\times n}$
    \Ensure $C\in\mathbb{R}^{m\times n}$
    \For{$i = 1,\dots,m$}
      \For{$j = 1,\dots,n$}
        \State $C[i,j] \gets 0$
        \For{$\ell = 1,\dots,k$}
          \State $C[i,j] \gets C[i,j] + A[i,\ell]\times B[\ell,j]$
        \EndFor
      \EndFor
    \EndFor
  \end{algorithmic}
\end{algorithm}

\begin{algorithm}
  \caption{Tiled Matrix Multiply Schedule}\label{alg:tiledMM}
  \begin{algorithmic}[1]
  \Require $M,N,K$ dimensions of the problem;\newline
            $M_T,N_T,K_T$ tile sizes in space and time
  \Ensure Partial results accumulated in $C$
  \Function{ScheduleTile}{$i,j,\ell$}
    \Comment{Compute one tile at CU indices $(i,j,\ell)$}
    \For{$m\gets 1$ to $M_{T}$}    \Comment{Across waves in a Block}
      \For{$n\gets 1$ to $N_{T}$}  \Comment{Across waves in a Block}
        \For{$k\gets 1$ to $K_{T}$}  \Comment{Across time in a wave}
          \State $C[i][j][m][n] \;\mathrel{+}=$
          \Statex[2] \quad $A[i][\ell][m][k]\;\times\,B[j][\ell][n][k]$
        \EndFor
      \EndFor
    \EndFor
  \EndFunction
  \For{$i\gets 1$ to $\lceil M/M_T\rceil$}    \Comment{Over CUs (space) in $M$}
    \For{$j\gets 1$ to $\lceil N/N_T\rceil$}  \Comment{Over CUs (space) in $N$}
      \For{$\ell\gets 1$ to $\lceil K/K_T\rceil$}  \Comment{Over time tiles}
        \State \Call{ScheduleTile}{$i,j,\ell$}
      \EndFor
    \EndFor
  \EndFor
  \end{algorithmic}
\end{algorithm}

One such example of how the output tile size impacts parallelism is the case where the matrix sizes are $M = 256$, $N=256$ and $K=8192$. A tile size of $(M_T\times N_T\times K_T) = (16\times16\times256)$ results in 256 output tiles ($\frac{M}{M_T} \times \frac{N}{N_T} = \frac{256}{16} \times \frac{256}{16}$, which
are parallelized over a GPU with 256 compute units (CUs), in a typical case each CU is responsible for producing one tile of the output matrix~\cite{Osama:2023:SWP}. Consider if we had chosen a tile size of $(M_T \times N_T \times K_T) = (256\times256\times128)$, results in a single output tile ($\frac{M}{M_T} \times \frac{N}{N_T} = \frac{256}{256} \times \frac{256}{256}$: The entire problem would run on a single CU resulting in underutilization of compute resources.

Separately optimizing for locality results in entries of a tile being reused for $(M_T\times N_T\times K_T)$ MAC operations. For example, a $16\times16\times256$ tile size is reused a total $65,536$ times whereas elements in a tile of $256\times256\times128$ are reused $8,388,608$ times\ganesh{I don't understand this. The numbers here are the number of elements in each tile; so why does it mean that they'll be reused that much?}. This is because tiling parameters change the amount of data loaded at a time into the lower levels of the memory hierarchy, i.e. one load to the lower level of caches can be reused for multiple matrix multiply operations. Tiling configurations are also recursive; most GPUs having some hardware predefined ``Matrix Instructions'' for a given data type, which will ingest a fixed size input tile as the base case of the recursion.

Contemporary approaches build GEMM libraries with autotuning as a solution that addresses the delicate balance of parallelism and locality. GEMM problems are benchmarked to determine the peak performance configurations. While autotuning results in the solution with 100\% efficiency in tuned cases, it takes $O(P \times M \times N \times K)$ for $P$ tiling configurations, with an additional complexity when adding different batch- or grouped GEMM-dimensions, and is not a scalable approach.

\subsection{Autotuning and Triton}
\label{sec:triton}
Triton is a domain-specific language and compiler framework designed for writing high-performance GPU kernels,
particularly those used in deep learning and tensor computation workloads~\cite{Tillet:2019:TAI}.
Triton aims to simplify GPU programming with a high-level Python interface that delivers performance comparable to hand-optimized kernels.
Triton kernels are written in Python and compiled Just-In-Time (JIT) to target specific GPU architectures.
This allows developers to focus on the algorithmic structure of kernels
while leaving many low-level optimization details to the compiler.

To achieve performance portability across different input sizes and hardware platforms,
Triton includes an autotuning system, a mechanism commonly found in
modern GPU programming frameworks such as TVM~\cite{Chen:2018:TVM},
TensorRT~\cite{NVIDIA:2024:TDG}, TensorFlow~\cite{Abadi:2016:TFS}, and
TorchInductor~\cite{PyTorch:2023:TIC}.
Autotuning refers to the process of automatically exploring a space of
implementation choices, such as tile sizes, memory tiling strategies,
loop unrolling factors, and warp scheduling parameters to discover configurations
that deliver optimal runtime performance for a specific target device and input shape.

In Triton, the autotuning mechanism is exposed through the \texttt{@triton.autotune} decorator.
Users define a set of candidate configurations using \texttt{triton.Config} objects,
where each configuration specifies compile-time parameters like tile sizes, number of warps/wavefronts, and pipeline stages.
Additionally, users provide a set of input-dependent tuning keys (e.g., tensor size or shape)
which determine when a new tuning search should be triggered.
When a Triton kernel is first invoked for a particular key,
the framework compiles all candidate configurations and benchmarks them on the target GPU
using an event-based mechanism to measure execution latency.
The configuration that yields the lowest execution time is selected as the optimal one
and is cached internally for later reuse in subsequent invocations with the same tuning key.

However, the autotuning process introduces several forms of overhead.
First, the initial tuning phase incurs compilation and measurement costs
which can be significant when the configuration space is large.
Second, the accuracy of the autotuner is bounded by the quality and completeness
of the candidate configuration space that is manually provided by the user.
Triton currently uses an exhaustive search over the user-defined configurations,
which limits scalability and adaptability to applications that have a wide variety of kernel shapes and sizes. 
Finally, the need to benchmark each configuration at runtime
makes this approach unsuitable and impractical for many applications, such as those with:
\begin{itemize}
  \item Dynamic tensor shapes and sizes that vary from run-to-run,
  \item Real-time constraints that demand low latency, such as online inference or control systems, and
  \item Applications with an already extremely high cost and power consumption such as LLM training runs.
\end{itemize}

The complex underlying relationship between GEMM kernel configurations, parallelism, locality, and performance has long been studied~\cite{BLIS1, BLIS2, BLIS3, BLIS4}. Many of these studies analytically derive the ideal tiling configurations based on architectural shapes and bottlenecks. We extend this concept to the highly parallel and complex GPU architectures with GEMM as an example. In the next two sections, we provide an analytical model for capturing these intricacies on the AMD Instinct\textsuperscript{TM} MI300X accelerator.

\subsection{Analytical Performance Models for GPU GEMM}
Recent GPU optimization systems increasingly employ analytical or semi-analytical models to reduce the cost of empirical autotuning. 
CUTLASS~\cite{cutlass} and cuBLAS~\cite{cublas}, for instance, use hand-engineered heuristics to assess the viability of threadblock tiles based on memory traffic, register pressure, and tensor-core utilization. These rules work well for CUDA templates but are closed sources and tightly coupled to NVIDIA-specific execution models which do not transfer cleanly to Triton. Likewise, Ansor and AutoTVM~\cite{ansor,autotvm} combine analytical reuse estimates with learned correction terms to prune large search spaces; however, their reliance on data-driven fitting makes them less suitable for dynamic-shape workloads or scenarios requiring deterministic, zero-sample prediction.

Several works also extend the roofline model to better capture locality in GPU kernels. 
DeLTA~\cite{delta}, for example, introduces a locality-aware roofline that predicts L1/L2 traffic via cache-blocking and reuse-distance analysis. Whereas DeLTA analyzes the performance of predefined kernel configurations, \emph{tritonBLAS} uses only calibrated hardware bandwidths and instruction latencies to directly rank and select GEMM tile shapes. This enables microsecond overhead, deterministic configuration selection within Triton, supporting dynamic shapes and eliminating the need for runtime autotuning.

\section{Design Goals}
The design goals of our analytical model and framework are:

\textit{\textbf{Achieve Near-Optimal Performance}}. The model must efficiently select GEMM parameters such that the GEMM operations achieve near-optimal performance for a wide range of matrix shapes and sizes.

\textit{\textbf{No-Tuning Required}}. The model must be designed as a general-purpose model that can be applied to any GEMM operation without the need for autotuning (exhaustive search) such as the one that exists within \framework{Triton}.

\textit{\textbf{Lightweight and fast}}. The model is meant to be used at \emph{runtime} to guide the tiling decisions of the GEMM operation without impacting performance.

\textit{\textbf{Deterministic}}. The model produces the same results for the same inputs, allowing for reproducibility and consistency in performance engineering.

\textit{\textbf{Architecture-Portable}}. The model must be agnostic to the GPU micro-architecture, meaning it can be applied to any GPU architecture with minimal adjustments and without the need for hardware access to tune. \muhammad{We will probably remove this.} \bryant{I think the fact that you can add new architecture support without access to the hardware is pretty central to the benefits.} \muhammad{I think the tricky part is how we show this without running stuff on MI350, which we cannot publish.} \alex{depending on how much additional time is available for edits, we could use some mi308, or cpx data} \ganesh{cpx/dpx/spx seems like an easy one to do}


\subsection{Non-Goals}
The model is not intended as a comprehensive performance model for GEMM latency prediction but rather to capture latency ``trends,'' enabling accurate comparative evaluation of GEMM parameter choices.

\textit{\textbf{Non-GEMM Workloads}}. This work is not intended for other GEMM-like algorithms such as various attention mechanisms, however we intend to explore that as part of our future works.

\textit{\textbf{Multi-GPU and Multi-Node}}. This work focuses on GEMM kernel implementations within a single GPU and is not yet intended for distributed system environments.


\section{tritonBLAS: Analytical Model}
We introduce \framework{tritonBLAS}, a triton based GEMM library including an analytical model that captures the interplay between hardware architecture and algorithmic structure by decomposing the GEMM operation into hierarchical memory and compute stages. Our model enables accurate selection of a tiling hierarchy that maximizes both parallelism and locality on the target GPU architecture. \framework{tritonBLAS} is built on the following key pillars described in detail in the subsections below: \muhammad{I'd like to keep this list to introduce the subsections.}

\begin{itemize}
    \item \textbf{Hierarchical Tiling Structure}
    \item \textbf{Quantifying Parallelism (Spatial Loop Unroll)}
    \item \textbf{Quantifying Locality}
    \item \textbf{Tradeoffs between Parallelism and Locality}
    \item \textbf{GEMM schedule Latency}
\end{itemize}





\subsection{Hierarchical Tiling Structure}

\begin{figure*}
  \includegraphics[width=\textwidth]{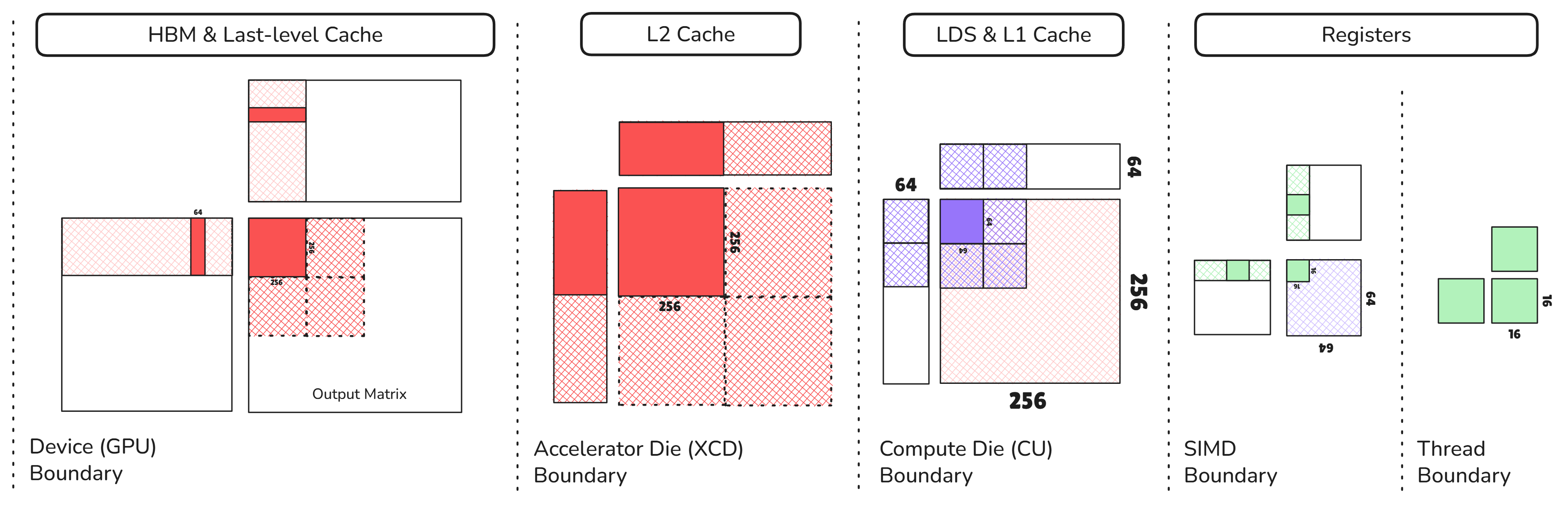}
  \centering
  \caption{Example of hierarchical tiling of output matrix for AMD's MI300X GPU (see Table~\ref{tab:mi300_taxonomy}, illustrating device, XCD, CU, SIMD and Thread breakdown of the tile).\label{fig:hierarchy} \ganesh{all numbers in the figs are illegible. Make everything the same size as the 256 in the 3rd picture and make it boldface} \muhammad{excalidraw doesn't export with font/text-styles, so I am using PNG instead of PDF/SVG here.}}
\end{figure*}

Our model captures the impact of tiling structures on parallelism and locality. Tiles are arranged in a hierarchy from the instruction-level ``atom'' of computation, to the overall system-level problem stored in main memory. The tiling hierarchy is defined as follows:

\begin{enumerate}
  \item Instruction Level Tile: The tile size consumed by the matrix instruction (\texttt{wmma}, \texttt{mfma}, etc.) in each SIMD,
  \item Warp/Wavefront (Register) Level Tile: A set of instruction tiles processed by a SIMD over time, 
  \item Workgroup/Thread block (Shared Memory) Level Tile: The Tile stored in software managed memory between SIMDs,
  \item Cache Tile: Above software managed memory are one or more hardware managed memories, each with some set of shared memory tiles belonging to their scope. This can be modeled as a simple hierarchy of loads with increasing scope sizes (based on how many shared memory tiles are being processed in that scope.
  \item Global Problem: The total overall problem unrolled over space (Compute Units) and time (when we run out of space).
\end{enumerate}

From the bottom of the tiling hierarchy to the top encompasses the entire problem size and how it is mapped to progressively larger sub-tiles. Levels of hierarchical tiling exist to maximize data locality and reuse, which then enables a higher total bandwidth, thus allowing modern GPUs to keep the increasingly data hungry computational units fed. For the MI300X GPU, we map the tiling hierarchy to the logical and physical paradigms in the hardware/programming model as shown in Table~\ref{tab:mi300_taxonomy}.

\begin{table}
  \begin{center}
    \resizebox{\columnwidth}{!}{%
      \begin{tabular}{c c c c}
      \toprule
      Memory Tiling Scope & Compute Scope & Physical Scope & Logical \\
      \midrule
      Instruction & Matrix Core & Matrix Core & Instruction \\
      \midrule
      Register & SIMD & SIMD & Wave \\
      \midrule
      Shared Memory & Compute Unit & Compute Unit & Workgroup \\
      \midrule
      L2 & Group of CU & XCD \tablefootnote{An XCD is a compute chiplet on the AMD Instinct\texttrademark~MI300X GPU, which contains some number of compute units that share an L2 cache.} & None \\
      \midrule
      LLC & All CUs & Device & None \\
      \bottomrule
      \end{tabular}
    }
    \end{center}

    \caption{The compute, memory, and logical scopes assigned to each level of tiles for an AMD Instinct\texttrademark~MI300X GPU~\cite{AMD:2023:cdna3}.}
    \label{tab:mi300_taxonomy}
\end{table}

We implement a Triton-based kernel which leverages each of these scopes by parameterizing hierarchical tiling configurations. This enables us to permute the tiling hierarchy through simple changes of kernel configurations. \framework{tritonBLAS} selects the kernel configurations by estimating the latency of the GEMM computation for each possible tiling hierarchy on a particular hardware architecture. To quantify the total system latency, we must consider the utilization of both the parallel compute and data movement resources.

\subsection{Quantifying Parallelism (Spatial Loop Unroll)}

In modern compute accelerators, parallelism is one key principle that results in performance. We use many different compute resources to parallelize computation at different levels of the hierarchy. We will consider three levels of parallelism: matrix/tensor core-level parallelism~\cite{AMD:2022:AMC,NVIDIA-Corporation:2019:TFO}, parallelism within a compute unit, and parallelism across compute units. Our model captures and quantifies the performance impact of each of these types of parallelism with well-defined deterministic expressions that do not require heuristic tuning. 

To formulate this expression, we use the latency of processing the \textit{lowest} level tile which is determined by how many operations are parallelized across the matrix/tensor cores. In most cases  a programmer has very little control over matrix/tensor core parallelism: it is defined by the architects of the GPU as the latency of the matrix instruction (e.g. \texttt{wmma} or \texttt{mfma} instructions~\cite{AMD:2025:AIM,NVIDIA-Corporation:2025:CCP}). This means that we can assume the latency is a constant, resulting in a consistent throughput of matrix instructions with dimensions ${MI}_M,{MI}_N,{MI}_K$ in some fixed amount of time $L_{MI}$ which can be used to parameterize all other computations.

\begin{algorithm}
  \caption{Calculate the Compute Latency of a Shared Memory Tile~\label{alg:compute_matrix_instruction_latency}}
  \begin{algorithmic}[1]
    \Require $MI_M, MI_N, MI_K$ \Comment{Dimensions of Matrix/Tensor Instruction}
    \Require $MT_M, MT_N, MT_K$ \Comment{Dimensions of Shared Memory Tile}
    \Require $L_{MI}$  \Comment{Latency of a matrix instruction}
    \Ensure $N_{MI}$\Comment{total Matrix Instructions per Shared Memory Tile}
    \Ensure $L_{MT}$\Comment{compute latency per Shared Memory Tile}
    \State $N_{MI,M} \gets \bigl\lceil \frac{MT_M}{\mathrm{MI}_M}\bigr\rceil$
    \State $N_{MI,N} \gets \bigl\lceil \frac{MT_N}{\mathrm{MI}_N}\bigr\rceil$
    \State $N_{MI,K} \gets \bigl\lceil \frac{MT_K}{\mathrm{MI}_K}\bigr\rceil$
    \State $N_{MI}   \gets N_{MI,M}\times N_{MI,N}\times N_{MI,K}$
    \State $L_{MT}  \gets L_{MI}\times N_{MI}$
  \end{algorithmic}
\end{algorithm}

Moving up the tiling hierarchy, we quantify the latency of the wavefront or register level tile which is parallelized over the set of SIMDs belonging to a compute unit. This parallelism means that the exposed latency for the shared memory or workgroup tile is exposed latency per register tile multiplied by the number of register tiles processed over time by a SIMD. Assuming the register tiles are processed in a way that results in a high utilization of the matrix multiplication acceleration unit, we can simplify the compute latency of a shared memory tile computation to be in terms of the dimensions and latency of the matrix instruction using Algorithm~\ref{alg:compute_matrix_instruction_latency}.


\begin{figure}
  \includegraphics[width=0.7\columnwidth]{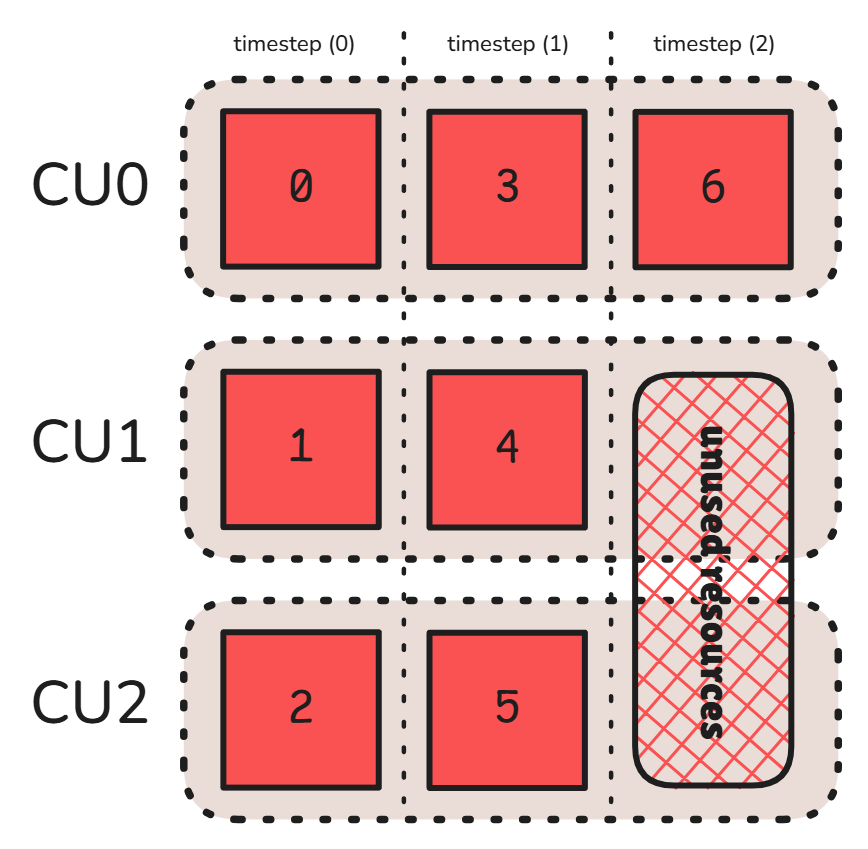}
  \centering
  \caption{Example of occupancy and CUs\label{fig:occupancy} across time (timestep$_0$, timestep$_1$ and timestep$_2$) and available CUs ($\text{CU}_0, \text{CU}_1, \text{CU}_2$). \ganesh{this figure is needlessly large for a relatively simple concept}} \muhammad{scaled it, I will rearrange it after the paper is completely done.}
\end{figure}

The last level of parallelism that \framework{tritonBLAS} quantifies is the utilization of the compute units in the entire system, or occupancy~\cite{Osama:2023:SWP}. To calculate occupancy, we first calculate using the dimension on which the problem is parallelized. Typical GPU GEMM implementations will map the $M$ and $N$ dimension across compute units in an output stationary dataflow. In this case, we parallelize workgroup tile dimension tensors in the $M$ and $N$ dimensions across space, and $K$ dimension processed sequentially processed over time. The occupancy is therefore determined by both the shape of the workgroup tile and the shape of the input problem in the $M$ and $N$ dimensions. Figure~\ref{fig:occupancy} shows, tiles parallelized across the CUs of a device with the occupancy of the total number of output tiles divided by the number of compute units. Figure~\ref{fig:occupancy} shows a key observation; the final timestep is the only ``under occupied'' timestep. This means that we can assume full occupancy (100\% of CUs utilized) for all but the last timestep (also known as tail occupancy or the wave quantization problem~\cite{Osama:2023:SWP}) to capture a complete view of occupancy of the entire problem. Algorithm~\ref{alg:occupancy} shows an algorithm for computing this tail latency.

\begin{algorithm}
  \caption{Compute Active Compute Units in last timestep.~\label{alg:occupancy}}
  \begin{algorithmic}[1]
      \Require $M, N, K$                           \Comment{Problem dimensions}
      \Require $MT_{M}, MT_{N}, MT_{K}$             \Comment{Macro‑tile sizes}
      \Require $N_{\mathit{CU}}$                    \Comment{Total number of CUs on hardware}
      \Ensure $active\_cu$,$\omega$                          \Comment{\# CUs active for last wave, \# Waves}
      \State $n_{MT,M} \gets \bigl\lceil \frac{M}{MT_{M}}\bigr\rceil$
      \State $n_{MT,N} \gets \bigl\lceil \frac{N}{MT_{N}}\bigr\rceil$
      \State $T_{\mathrm{out}} \gets n_{MT,M} \times n_{MT,N}$
      \State $\omega \gets \bigl\lceil \frac{T_{\mathrm{out}}}{N_{\mathit{CU}}}\bigr\rceil$
      \State $active\_cu \gets T_{out} \% N_{CU}$
      \State \Return $active\_cu$, $\omega$
  \end{algorithmic}
\end{algorithm}

The degree of parallelism in GEMM across compute units influences not only the number of compute units utilized but also the availability of load and store units. This is particularly critical for small GEMM problems where a careful balance must be struck: issuing enough workgroups to sustain load/store throughput without fragmenting tiles to the point of compromising locality. An overly aggressive approach can undermine performance due to poor caching, despite increased parallelism. The following subsections detail how \framework{tritonBLAS} captures this interplay between parallelism and locality.

\subsection{Quantifying Locality}
In \framework{tritonBLAS} we define two types of data locality: software managed and hardware managed. Modern GPU architectures support both paradigms using programmer-controlled memory (software managed) and caches (hardware managed). An example of software managed memory includes shared memory, where data movement is explicitly specified. In contrast, hardware managed locality relies on the hardware to transparently manage data placement and movement, as seen in caches. In the workgroup tile for example, the $M \times K$ tile is reused $N$ times and the $N \times K$ is reused $M$ times. This enables the programmer to precisely control data reuse within a SIMD and data sharing across SIMDs, thanks to the explicit management of shared memory.

The more implicit ``cache tile'' dimensions may be computed using the number of workgroup tiles 
that have access to a given cache. For example, if there are $16$ compute units each processing their own workgroup tile, the area of the cache tile will be $16$, and the options for arrangement of that cache tile are factorizations of $16$. 
Within \framework{tritonBLAS} we also predict the shape of the tile shared in a cache in a way that maximizes data reuse between the occupied compute units. Algorithm~\ref{alg:cache_factorization} shows an example of how to compute possible cache tile dimensions given a number of compute units belonging to a particular cache scope. Most often, the largest square formed out of compute units sharing the cache for a certain problem ensures the best data reuse, so in cases where we require a default, such as tile computation, we can often use $floor(sqrt(N_{\mathrm{CU\_Cache}}))$ as a default factorization.

\begin{algorithm}
  \caption{Estimate Cache Hit Rate from tile dimensions.~\label{alg:hitrate}}
  \begin{algorithmic}[1]
    \Require Cache-tile dimensions $m_t, n_t$; 
    \Require Workgroup tile dims.\ $M_T, N_T, K_T$
    \Ensure Hit rate $h$
    \State $U \gets (m_t\,M_T + n_t\,N_T)\,K_T$ \Comment{Uncached reads}
    \State $R \gets (m_t\,n_t)\,(M_T + N_T)\,K_T$ \Comment{Total reads}
    \State $h \gets 1 - \dfrac{U}{R}$ \Comment{Hit rate}
    \Ensure $0 \le h \le 1$
    \State \Return $h$
  \end{algorithmic}
\end{algorithm}

Algorithm~\ref{alg:hitrate} estimates the cache hit rate required for modeling memory latency in our analytical framework. The formulation assumes that higher-level caches interact with data in units defined by the tiles of the layer below, a key principle of hierarchical tiling. Thus, uncached reads ($U$) scale with the number of distinct tiles accessed per group of thread blocks, while total reads ($R$) reflect the full volume of memory traffic from all thread blocks.


\begin{algorithm}
  \caption{Compute all tile-factorizations of $N_{\mathrm{CU\_Cache}}$.}
  \label{alg:cache_factorization}
  \begin{algorithmic}[1]
    \Require Number of compute units in cache scope: $N_{\mathrm{CU\_Cache}}$
    \Ensure A list $\mathcal{F}$ of all pairs $(f_m,f_n)$ such that $f_m \times f_n = N_{\mathrm{CU\_Cache}}$
    \State $\mathcal{F} \gets [\,]$ \Comment{Initialize empty list of factor pairs}
    \For{$i = 1$ \textbf{to} $N_{\mathrm{CU\_Cache}}$}
      \If{$N_{\mathrm{CU\_Cache}} \bmod i = 0$}
        \State $j \gets \dfrac{N_{\mathrm{CU\_Cache}}}{i}$
        \State Append $(\,i,\,j\,)$ to $\mathcal{F}$
      \EndIf
    \EndFor
    \State \Return $\mathcal{F}$
  \end{algorithmic}
\end{algorithm}

Our model incorporates cache capacity by comparing the tile-level working-set size at each K-step to the effective cache capacity and proportionally reducing the predicted hit rate whenever the footprint exceeds that limit. This captures the dominant capacity effects for GEMM, where reuse is governed mainly by tile-level temporal locality across K-steps. We do not model set associativity or replacement behavior, as these have limited influence on the large, regular access patterns in GEMM and would add significant complexity for only modest accuracy gains.


Using the hit rates for each level of the memory hierarchy, the memory latency for an example memory hierarchy with 2 caches before global memory can be computed using algorithm~\ref{alg:memory_latency}. 

\begin{algorithm}
  \caption{Calculate the Memory Latency of a Loop Iteration.~\label{alg:memory_latency}}
  \begin{algorithmic}[1]
    \Require Problem dims.\ $M,N,K$, tile dims.\ $M_T,N_T,K_T$
    \Require Hit rates $H_{1},H_{2}$ for Mem1 and Mem2
    \Require Number of CU Loads $Ld_{\mathrm{CU}}$, active CUs $C$, total CUs $C_{\max}$
    \Require Bandwidths $R_{L1},R_{1},R_{2},R_{\mathrm{MEM}}$ in elements per compute cycle
    \Require memory latency $L_{\mathrm{lat}}$
    \Ensure Per-CU memory‐bound latency $L_{\mathrm{mem}}$
    \State $L_{\mathrm{CU\_lat}} \gets \dfrac{Ld_{\mathrm{CU}}}{R_{L1}}$
    \State $T \gets Ld_{\mathrm{CU}}\times C$ \Comment{Total loads across active CUs}
    \State $L_{1} \gets \dfrac{T}{R_{1}}$ \Comment{Latency of loads through Mem1}
    \State $T_{2} \gets (1 - H_{1})\,T$ \Comment{Loads missing in Mem1}
    \State $L_{2} \gets \dfrac{T_{2}}{R_{2}}$ \Comment{Latency of loads through Mem2}
    \State $T_{\mathrm{M}} \gets (1 - H_{2})\,T_{2}$ \Comment{Loads missing in Mem2}
    \State $L_{\mathrm{MEM}} \gets \dfrac{T_{\mathrm{M}}}{R_{\mathrm{MEM}}} + L_{\mathrm{lat}}$ \Comment{Latency of memory loads}
    \State 
    \State $L_{\mathrm{mem}} \gets \max\bigl(
        L_{\mathrm{CU\_lat}},\,L_{1},\,L_{2},\,L_{\mathrm{MEM}}
    \bigr)$
    \State \Return $L_{\mathrm{mem}}$
  \end{algorithmic}
\end{algorithm}

\subsection{Tradeoffs between Parallelism and Locality}
The best tile for a problem is the one that achieves the lowest latency, however this does not necessarily correlate with the highest locality or parallelism. There are fine-grained trade-offs between parallelism and locality that determine the overall latency of a single loop iteration. Our analytical model quantifies the compute and memory latencies of one K-dimension iteration (over time). Due to software pipelining, the exposed computation latency is the maximum of the two latencies—data-movement/locality versus parallelism/occupancy—for a single iteration.

This means that, for each unique problem, there is a unique entry in the tradeoff space of parallelism and locality. The tradeoff can be conceptualized as a set of competing rooflines representing utilization of the different functional units (parallelism) and data busses (locality). Different tiling selections can be broken into categories based on which roofline their bottleneck belongs to for a particular problem. A coarse-grained set of bottlenecks that arise are:

\begin{itemize}
  \sloppy
  \item \textbf{Load/Store Issue Rate Bound}: There are not enough load/store units occupied to produce enough bandwidth.\\ \textbf{Solution:} Increase load/store unit occupancy.
  \item \textbf{Software Managed Memory Bandwidth Bound}: Making the most reuse of the Shared Memory, but are bound by loads from it.\\ \textbf{Solution:} Increase register tile size to maximize reuse of data loaded from software managed memory.
  \item \textbf{Cache Bandwidth Bound}: The hardware managed caches are bounding the computation due to limited bandwidth.\\ \textbf{Solution:} Increase data reuse in software managed memory.
  \item \textbf{Under-Occupied Compute Bound}: The utilization of the matrix instructions is high in the occupied compute units, but not enough compute units are occupied to achieve max throughput.\\ \textbf{Solution:} Unroll K-loop or change tile size to achieve higher occupancy.
  \item \textbf{Max Parallelism Compute Bound}: All compute units are occupied, and are achieving a high utilization of their matrix instruction \alex{should this be "matrix cores"?} \ryan{I think we were trying to be more general than matrix core, which is an AMD specific term.} due to a non memory bound software pipeline. In this state, we achieve the maximum possible performance.
\end{itemize}

Once certain factors are fully optimized, such as achieving compute-bound execution with full occupancy, further optimization may no longer be possible. However, many of these objectives are inherently conflicting: For instance increasing the load/store issue rate often requires greater parallelization across cores, which in turn necessitates smaller tile sizes. Smaller tiles, however, reduce data reuse in the cache hierarchy since reuse is closely tied to tile dimensions. The analytical model navigates these trade-offs by estimating total computational latency based on first-principles analysis of the underlying bottlenecks.


\subsection{Quantifying GEMM Latency}

Lastly, all of these individual parameters are aggregated into a final ``latency'' as shown in Algorithm~\ref{alg:gemm_total_latency} which is used as a quantitative metric for the quality of a tiling solution for a problem. We must take into account that the pipeline is limited by the latency of the memory loads and compute instructions, whichever is larger, and that additional pipeline bubbles may exist. We call the pipeline bubble at the beginning (loads but no compute) the ``prologue'' and the pipeline bubble at the end (writes and no compute) the ``epilogue''. The prologue scales
the same as the load stage of the software pipeline, but the store stage scales only in the output tile dimensions $({MT}_M,{MT}_N)$
instead of the input tile dimensions $({MT}_M,{MT}_N,{MT}_K)$. Unlike the pipeline latency which will be incurred for each iteration in $\frac{K}{{MT}_K}$, the prologue and epilogue happen once per output tile.

\begin{algorithm}
  \caption{Compute the latency to complete an output tile \label{alg:tile_latency}}
  \begin{algorithmic}[1]
    \Require hardware parameters $H$, problem dimensions $M,N,K$, tile sizes $m_t,n_t,k_t$
    \Require $L_{mem} $ previously computed memory latency of tile loads
    \Require $L_{compute}$ previously computed compute latency of tile loads
    \Require $a\gets$ \# of active CUs
    \Require Bandwidths $R_{L1},R_{1},R_{2},R_{\mathrm{MEM}}$ in elements per compute cycle
    \Ensure tile latency $L_{\mathrm{tile}}$
    \State $L_{\mathrm{prologue}}\gets L_{\mathrm{mem}}$
    \State $L_{\mathrm{epilogue}}\gets \dfrac{a\cdot m_t\cdot n_t}{H.R_{\mathrm{mem}}}$
    \State $L_{\mathrm{loopiter}}\gets \max\bigl(L_{\mathrm{compute}},\,L_{\mathrm{mem}}\bigr)$
    \State $I\gets (\lceil K / k_t \rceil - 1)$ \Comment{\# of loop iterations}
    \State $L_{\mathrm{tile}}\gets L_{\mathrm{prologue}} + L_{\mathrm{epilogue}} + (L_{\mathrm{loopiter}} \times I)$
    \State
    \Return $L_{\mathrm{tile}}$
  \end{algorithmic}
\end{algorithm}


However, as shown in Figure~\ref{fig:occupancy}, a computation may be exposed to multiple timesteps if the number of output tiles exceeds the number of compute units. This means the number of timesteps are directly related to the number of CUs in an architecture. Each timestep will have the latency of a single output tile, meaning we can use our previously mentioned output tile latency (assuming it produces $N_{CU}$ output tiles) and the number of timesteps to compute the total GEMM operation latency using the computation shown in algorithm~\ref{alg:gemm_total_latency}. 
  
\begin{algorithm}
  \caption{Compute Total GEMM Latency.~\label{alg:gemm_total_latency} 
  }
  \begin{algorithmic}[1]
    \Require hardware $H$
    \Require problem dims $M,N,K$, 
    \Require tile dims $m_t,n_t,k_t$
    \Ensure total latency $L_{\mathrm{total}}$
    \State $n_{m}\gets \lceil M / m_t\rceil$, \quad $n_{n}\gets \lceil N / n_t\rceil$
    \State $N_{\mathrm{waves}}\gets \lceil (n_{m}\times n_{n}) / H.N_{\mathrm{CU}}\rceil$
    \State $L_{\mathrm{total}}\gets N_{\mathrm{waves}}\times L_{\mathrm{tile}}$
    \State
    \Return $L_{\mathrm{total}}$
  \end{algorithmic}
\end{algorithm}

\section{Performance Evaluation}
We present results on an AMD Instinct\texttrademark~MI300X GPU using Triton version 3.4.0 using \texttt{rocm/pytorch:rocm6.4.1} docker container. The results are presented as follows:
\begin{itemize}
    \item Accuracy of \framework{tritonBLAS} versus Triton's \texttt{@triton.autotune} exhaustive search,
    \item Analysis of the overhead of model prediction time for \framework{tritonBLAS} and exhaustive search for Triton.
    \item General performance of \framework{tritonBLAS} compared to vendor's peak GEMM implementation (using PyTorch.)
    \item Performance on key matrix sizes derived from Llama3 model.
\end{itemize}


In our examples, we choose the \texttt{float16} (FP16) data type with both matrices being contiguous in K in memory.

\subsection{Accuracy of \framework{tritonBLAS} vs. \texttt{@triton.autotune}}
We compared \framework{tritonBLAS}'s tile selection against the tiling selection achievable through an exhaustive search over all tiling combinations. In our experiment we used 150,000 random problem sizes whose dimensions are multiples of 128 less than 8193. We measure the performance of each set of tiling parameters exhaustively. We report the 
\textbf{selection efficiency}. The selection efficiency is defined as the percent of achieved performance by tritonBLAS relative to the maximum achieved performance in the exhaustive search of all options.
In this experiment \framework{tritonBLAS} achieves a 94.7\% selection efficiency relative to the exhaustive search. Without
the overhead of autotuning, \framework{tritonBLAS} can achieve near-optimal performance of the exhaustive search based solution.

\begin{figure}[h]
\centering
  \includegraphics[width=\columnwidth]{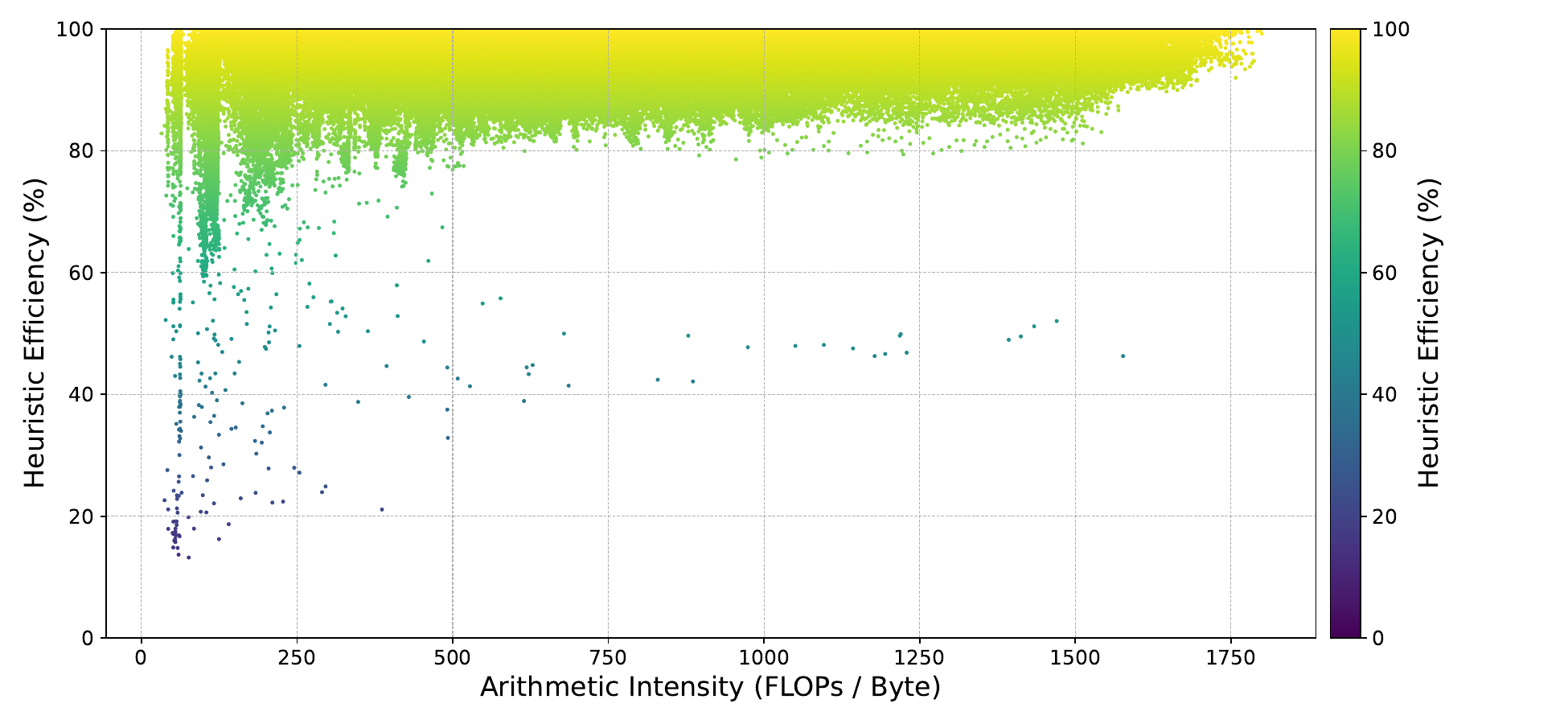}
  \caption{Efficiency of \framework{tritonBLAS} relative to Triton's autotune exhaustive search across 150,000 sizes.}
  \label{fig:eff}
\end{figure}

Figure~\ref{fig:eff} visualizes the achieved efficiency achieved per problem over the 150,000 problem shapes and sizes. The x-axis represents arithmetic intensity (flops/byte) of a problem, the y-axis represents efficiency achieved by the model, with 1.0 being best kernel predicted. Most data points are densely clustered near the top of the graph, indicating high heuristic efficiency across a wide range of arithmetic intensities, with more variation at lower intensities (latency-bound problems).

\subsection{Overhead of using \framework{tritonBLAS}}
\label{sec:overhead}


\framework{tritonBLAS} is capable of significantly faster solution selection than the existing tuning-driven selection process. This is enabled by \framework{tritonBLAS}'s non-iterative math on the problem and tile dimensions, only looping over the number of parameters.  Autotuning (exhaustive search) based approaches on 
the other hand will run the kernel for each parameter on each problem size, resulting in a much larger selection overhead. Solution selection in tritonBLAS adds a static cost of estimating latency for every candidate tile but removes the cost of exhaustively building and benchmarking each of the kernels resulting in a significant savings in the latency of new problem shapes. In other words, \framework{tritonBLAS} has a purely linear overhead in $P$, while exhaustive search incurs an additional multiplicative factor of $M N K$. This problem becomes worse when we account for batched or grouped GEMM, where the cost/complexity for contemporary approach scales to $\Theta(P\,M\,N\,K\,B)$, where $B$ is the number of batches.

Table~\ref{tab:overhead} demonstrates a striking advantage of \framework{tritonBLAS} over traditional Triton autotuning in terms of tile selection time. While autotuning incurs a cost proportional to the number of tile candidates (e.g., over 11 seconds for 75 tiles), the heuristic-based selection reduces this to the order of microseconds, representing an improvement of five to six orders of magnitude. In both instances, the results of parameter selection can be cached resulting in a subsequent selection for the same problem taking on the order of 1s of microseconds.
tritonBLAS performs analytical tile selection in 50–80 us, independent of GEMM size, whereas Triton’s autotuning requires compiling and benchmarking every candidate configuration, taking 10–50 seconds for the same 75 configurations (Table II). This means tritonBLAS typically selects a near-optimal kernel more than five orders of magnitude faster than autotuning. This makes tritonBLAS far more suitable for dynamic workloads, large search spaces, or environments where tuning cost dominates runtime.

\begin{table}
\centering
\resizebox{\columnwidth}{!}{%
  \begin{tabular}{lll}
  \toprule
  \textbf{Problem Size} & \textbf{Triton Autotuning (s)} & \textbf{tritonBLAS (s)} \\
  ($M \times N \times K$) & $\mathcal{O}(TMNK)$ & $\mathcal{O}(T)$ \\
  \midrule
  $512\times512\times512$       & 11.965 & 0.000070   \\
  $1024\times1024\times1024$    & 11.928 & 0.000057   \\
  $2048\times2048\times2048$    & 12.303 & 0.000073   \\
  $4096\times4096\times4096$    & 13.537 & 0.000055   \\
  $8192\times8192\times8192$    & 48.087 & 0.000075   \\
  $16384\times16384\times16384$ & 1383.594 & 0.000054 \\
  \bottomrule
  \end{tabular}
}
\caption{Comparison of selection time (in seconds) of Triton Autotuning vs \framework{tritonBLAS} Heuristic and Tile Selection with $75$ configurations searched. Shows that \framework{tritonBLAS} does not scale with the problem sizes and is orders of magnitude faster selection than tuning-based approach.} 
\label{tab:overhead}
\end{table}

\subsection{General Performance of \framework{tritonBLAS} in Triton}

\begin{figure*}[t]
  \centering

  \begin{subfigure}{0.48\textwidth}
    \centering
    \includegraphics[width=\linewidth]{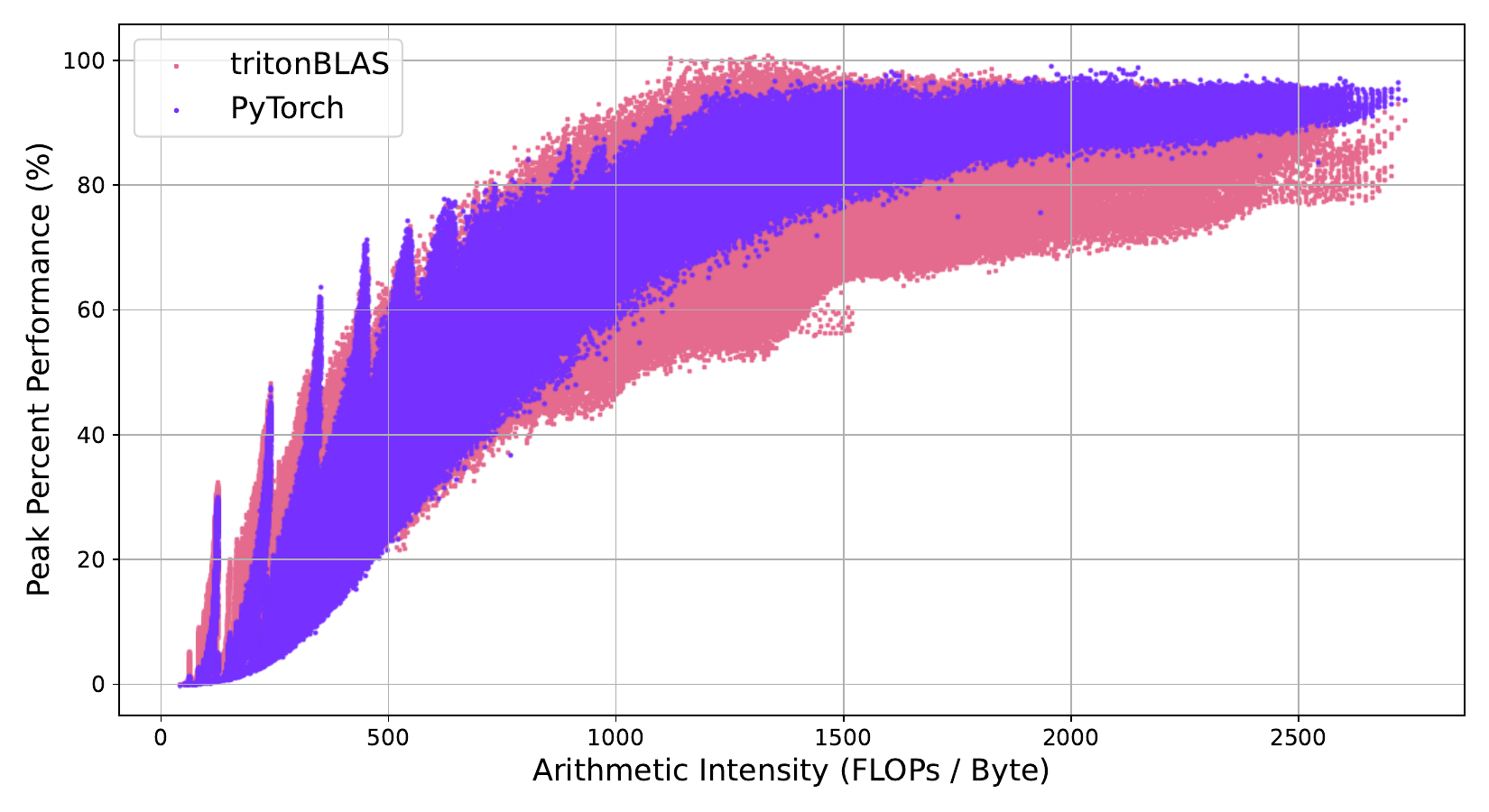}
    \caption{Peak percent performance of \framework{tritonBLAS} and \texttt{torch.matmul()} on AMD's Instinct\texttrademark~MI300X. Peak percent throughput is normalized to maximum achievable performance as highlighted by Ben Sander~\cite{Sander:2025:UPM,Sander:2025:MMA}. We show how closely the \framework{tritonBLAS} implementation approaches the performance of the optimized torch's tuned and optimized implementation.}
    \label{fig:vspytorch}
  \end{subfigure}
  \hfill
  \begin{subfigure}{0.48\textwidth}
    \centering
    \includegraphics[width=1.1\linewidth]{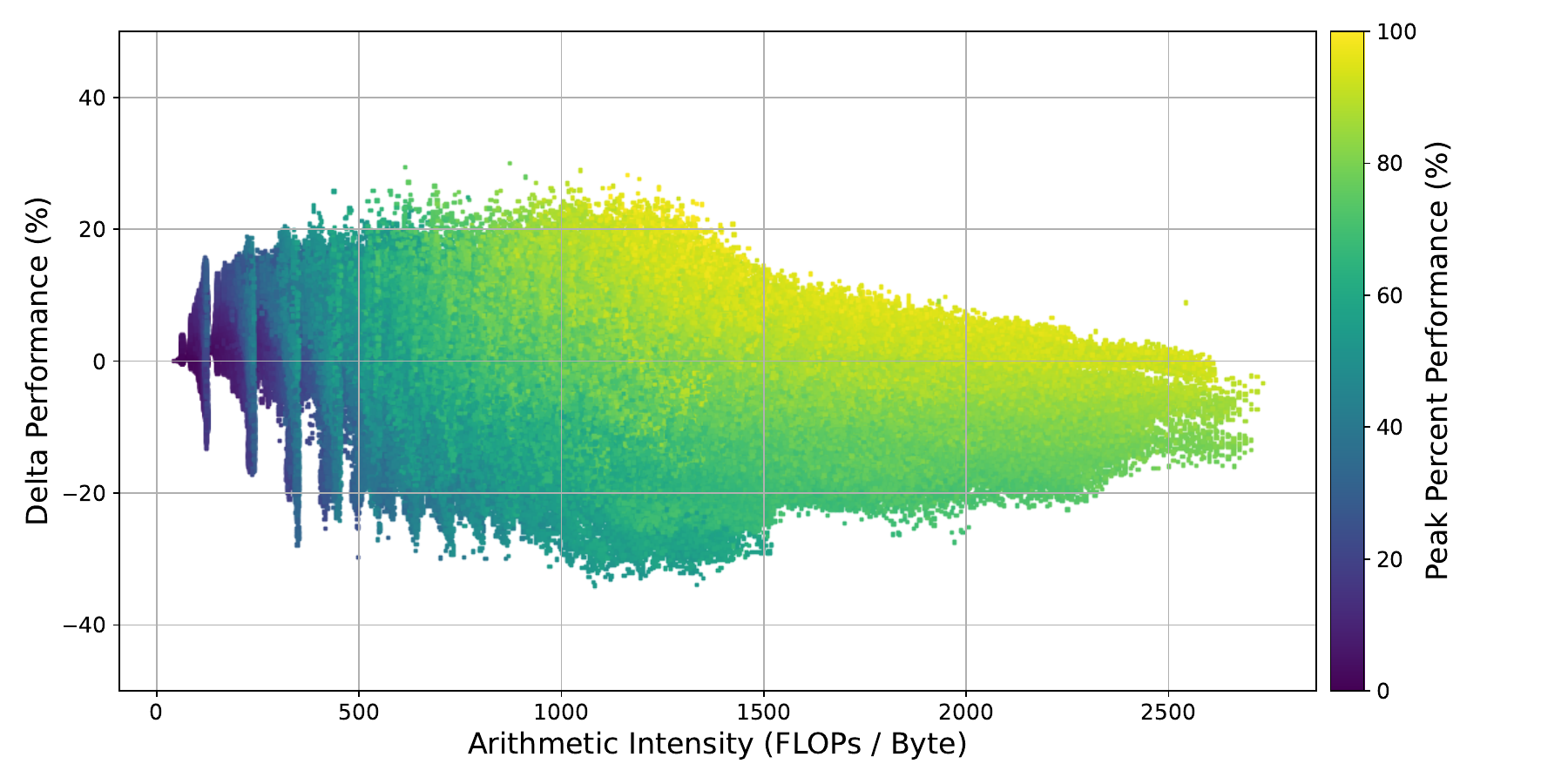}
    \caption{Delta percent between \texttt{torch.matmul()} optimized GEMM implementation and \framework{tritonBLAS}. Tritonblas performs within 20\% of vendor libraries called by \texttt{torch.matmul()}\\\\\\}
    \label{fig:delta_performance}
  \end{subfigure}

  \vspace{0.5cm}

  \begin{subfigure}{0.48\textwidth}
    \centering
    \includegraphics[width=\linewidth]{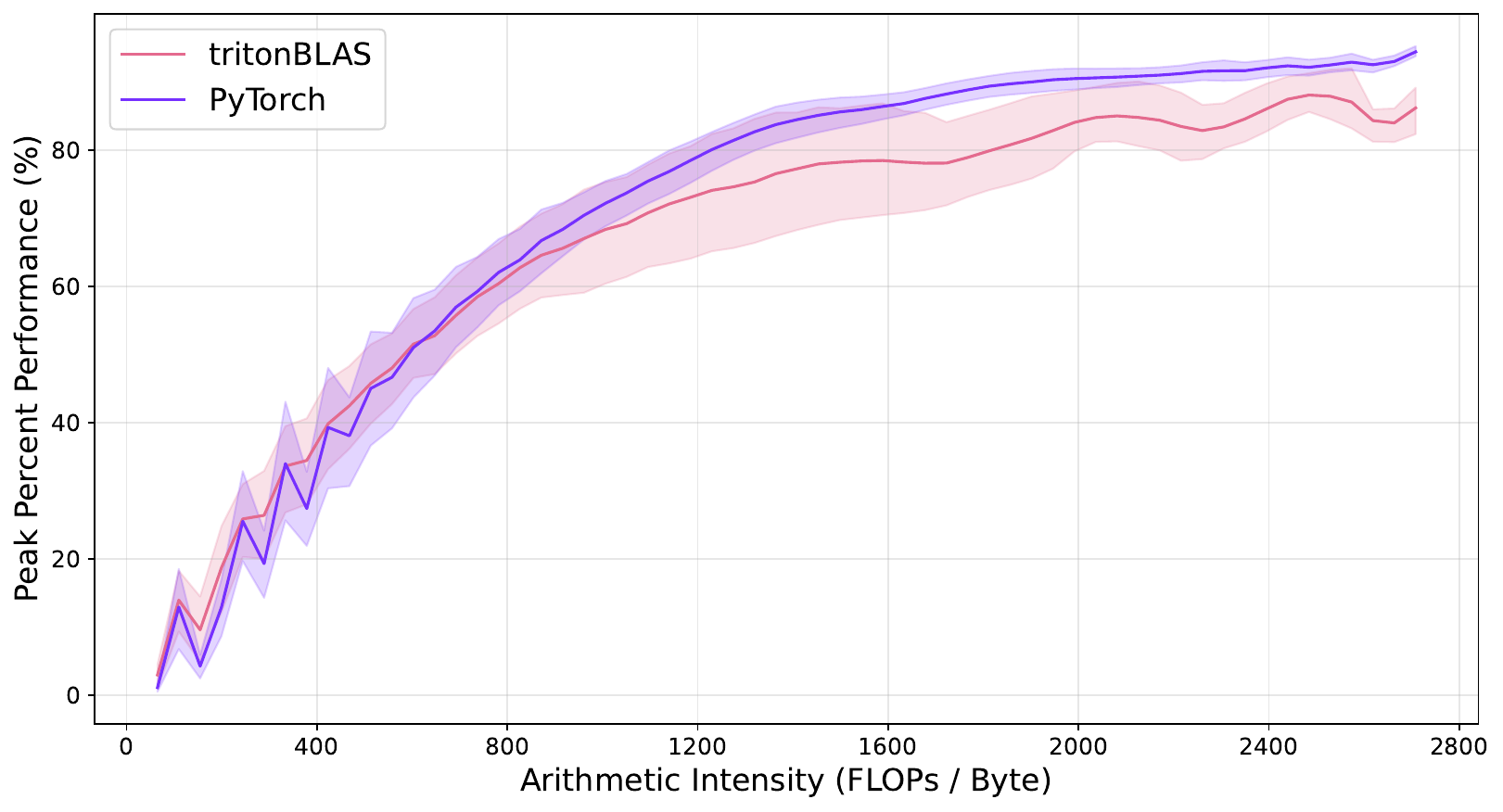}
    \caption{Percent of Peak as a function of Arithmetic Intensity of \framework{tritonBLAS} and \texttt{torch.matmul()} on AMD's Instinct\texttrademark~MI300X. \label{efficiency_spread}}
    \label{fig:efficiency_spread}
  \end{subfigure}
  \hfill
  \begin{subfigure}{0.48\textwidth}
    \centering
    \includegraphics[width=\linewidth]{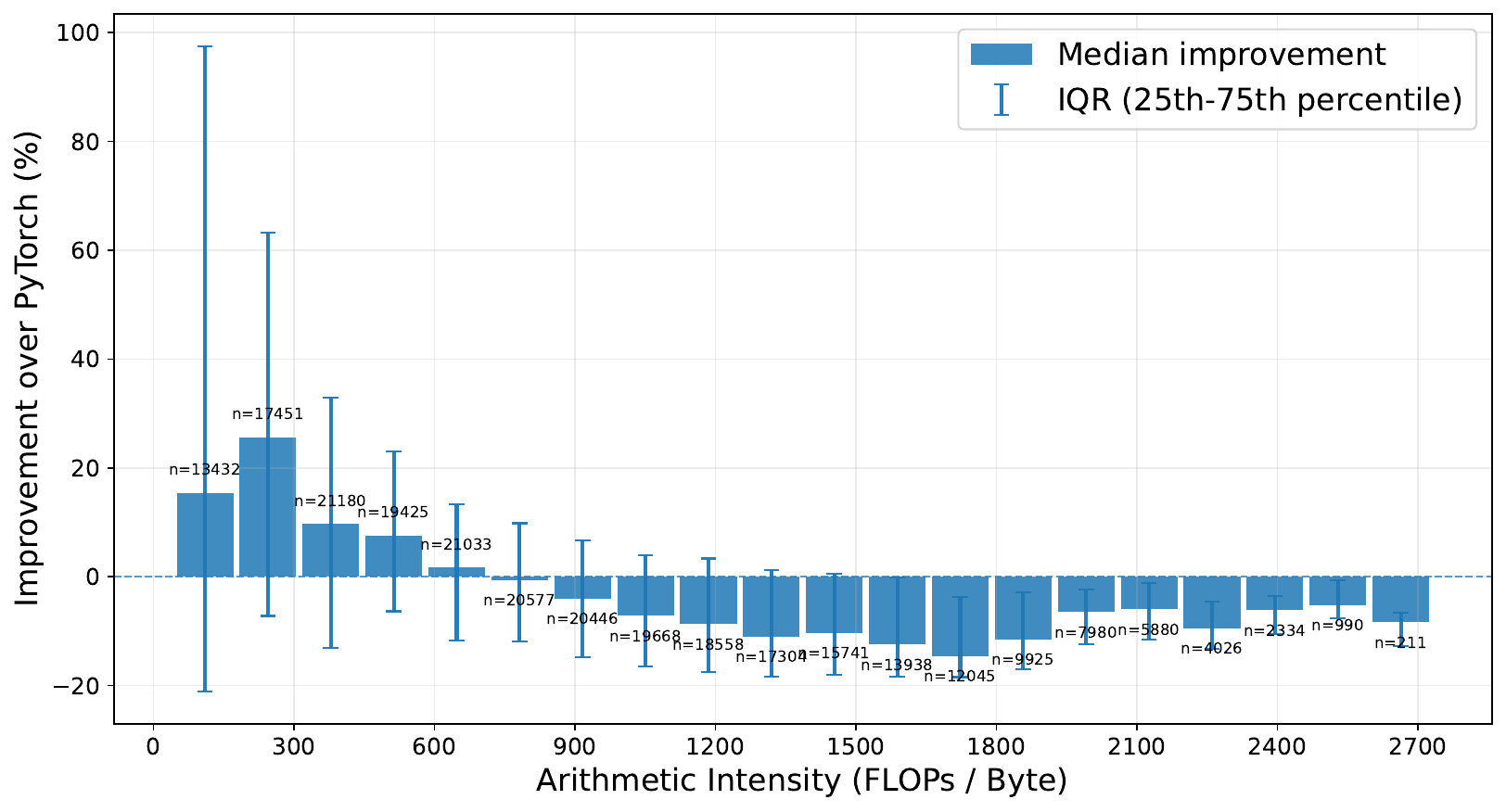}
    \caption{Binned peak percent performance of \framework{tritonBLAS} and \texttt{torch.matmul()} on AMD's Instinct\texttrademark~MI300X. \label{benchmark_complexity_histogram}}
    \label{fig:benchmark_complexity_histogram}
  \end{subfigure}

  \caption{Comparison between \framework{tritonBLAS} and \texttt{torch.matmul()} on MI300X, showing peak percent performance (top-left), delta percent (top-right), performance vs. arithmetic intensity (bottom-left), and binned peak percent performance (bottom-right).}
  \label{fig:combined_performance_fullwidth}
\end{figure*}

\begin{figure}[h!]
\centering
  \includegraphics[width=\columnwidth]{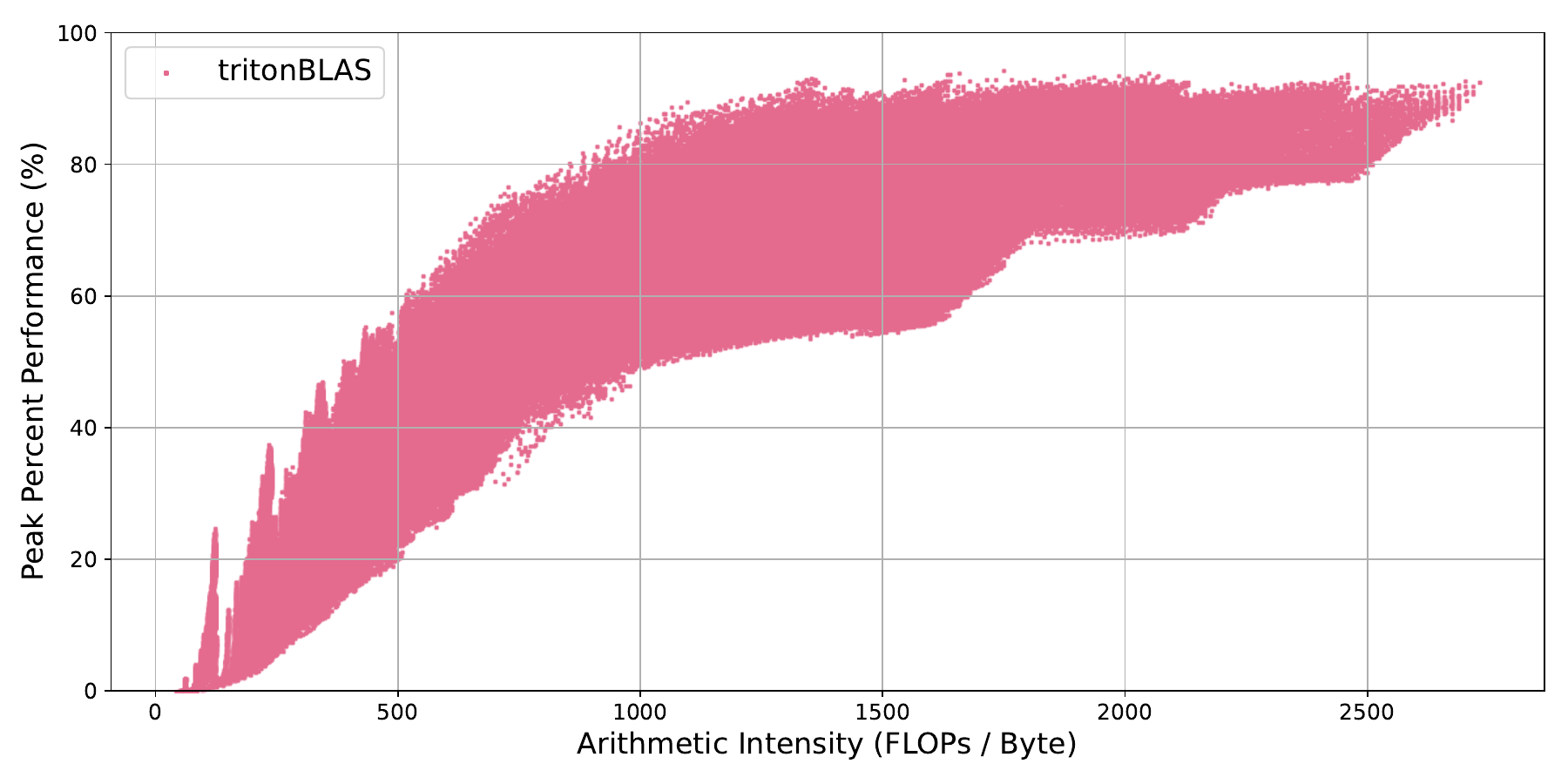}
  \caption{\framework{tritonBLAS} performance characteristics on AMD's Instinct\texttrademark~MI350X. The \framework{tritonBLAS} framework demonstrates strong performance on AMD Instinct™ MI350X, requiring only minor constant adjustments from the MI300X implementation. \label{fig:mi350_perf}}
\end{figure}

PyTorch matrix multiply also provides a python accessible method to perform matrix multiply on GPU using
vendor libraries which, in the case of MI300X, are autotuned to achieve peak performance on the hardware.
The results in Figure~\ref{fig:vspytorch} show a comparison of the performance of our \framework{tritonBLAS} implementation with PyTorch's matrix multiplication function. \framework{tritonBLAS} uses techniques such as Stream-K~\cite{Osama:2023:SWP}, to manage wave quantization to get better GPU occupancy for all GEMM shapes and sizes. On average, \framework{tritonBLAS} is 3\% better than PyTorch's matrix multiplication.

Figure~\ref{fig:vspytorch} also shows some performance gap from PyTorch, despite the efficiency of selection shown in Figure~\ref{fig:eff}. While some of this gap can be attributed to the ~5\% average gap in selection efficiency we also find a significant portion comes from tile quantization; libraries such as the one used by PyTorch, tiling parameters are available in increments of matrix dimensions. Triton however, does not support tiles which do not have power of 2 tile dimensions, limiting options for managing tile quantization~\cite{Tillet:2019:TAI,Zhou:2025:LL}. 



\subsection{\framework{tritonBLAS} on Key Llama3 Matrices}

\begin{figure*}
  \centering
  \includegraphics[width=\textwidth]{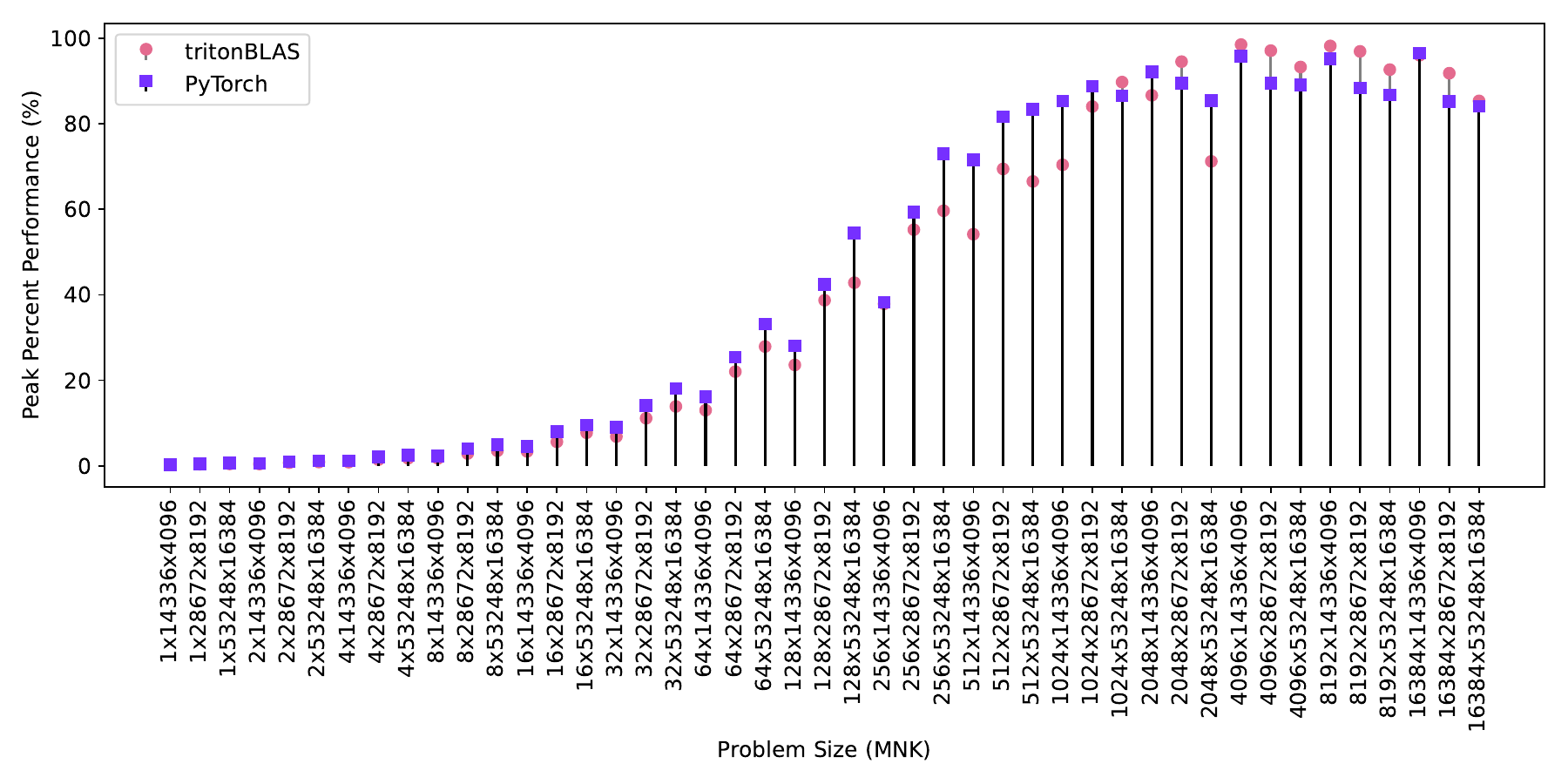}
  \caption{Delta percent between \texttt{torch.matmul()} and \framework{tritonBLAS} on key Llama3~\cite{Meta:2024:IML} matrix sizes. This plot shows how \framework{tritonBLAS}'s selection is able to match the performance of optimized GEMM implementation.}
  \label{fig:e2e_llama3}
\end{figure*}

Figure~\ref{fig:e2e_llama3} illustrates the performance comparison between \framework{tritonBLAS} and \texttt{torch.matmul()} across a range of problem sizes on the MI300X architecture. While Figure~\ref{fig:vspytorch} highlights overall performance trends, this plot emphasizes that achieving high performance on key matrix sizes is equally important. The sizes shown here are derived from the Llama 3 model workload~\cite{Meta:2024:IML}, making them representative of real inference demands. Our results show that \framework{tritonBLAS} competes closely with the peak performance of \texttt{torch.matmul()}, with a maximum observed speedup of 1.10$\times$ and outperforming PyTorch in 10 cases. Although \framework{tritonBLAS} is on average 13.9\% slower, it still delivers performance that is highly competitive with the best kernels available for this hardware.


\subsection{Architecture Portability}
Triton exposes a large configuration space with 50–150 valid tiles per GEMM depending on datatype. Architectural constraints create sharp performance cliffs and multiple disjoint optima, making the space unfriendly to random search methods. We display that our dataset is representative of many tradeoffs and arithmetic intensities without bias towards a particular intensity in figure \ref{benchmark_complexity_histogram} and show the average performance in bins of intensity in both \ref{benchmark_complexity_histogram} and \ref{efficiency_spread}

The analytical model is architecture-portable as it depends only on a small set of calibrated hardware parameters: cache and memory bandwidths, access latencies, and MFMA/Tensor Core instruction shapes. Updating these values is sufficient to retarget the model to a new GPU. As shown in Fig. \ref{fig:mi350_perf}, applying the same model to AMD’s MI350X (with only these parameters changed) yields performance trends consistent with MI300X, demonstrating straightforward cross-architecture transfer. MI300X remains our primary evaluation platform due to resource constraints, but the MI350X results confirm that no additional tuning or model changes are required.

\section{Conclusion and Future Work}
\framework{tritonBLAS} introduces a deterministic, analytical framework for GEMM kernel parameter selection that eliminates the need for runtime autotuning. By modeling the interplay between GPU architecture, tiling hierarchies, parallelism, and data locality, \framework{tritonBLAS} achieves near-optimal performance, 94.7\% of exhaustive search efficiency, while drastically reducing configuration overhead. The model’s architecture-agnostic design and reproducibility make it suitable for diverse workloads and hardware platforms. Performance evaluations demonstrate its scalability, low latency, and competitive throughput compared to vendor-optimized libraries. This work lays the foundation for future extensions to multi-GPU environments and broader algorithm classes, offering a practical and efficient alternative to empirical tuning in high-performance computing and machine learning applications.

\section{Acknowledgments}
The authors would like to thank Minsu Kim, Babak Poursartip, Vinayak Gokhale, Mike Schulte, Ralph Wittig, Brad Nemanich and Khasid Ali Khan for the their continuous feedback and support in the development of Origami and tritonBLAS. AMD, the AMD Arrow logo, AMD CDNA\texttrademark, AMD Instinct\texttrademark, AMD ROCm\texttrademark,
AMD Infinity Cache\texttrademark, AMD Infinity Fabric\texttrademark, and combinations
thereof are trademarks of Advanced Micro Devices, Inc\@. Other
product names used in this publication are for identification
purposes only and may be trademarks of their respective
companies.

\newpage
\bibliographystyle{IEEEtran}
\bibliography{references}


\end{document}